\documentstyle[12pt]{article}

\author{{\bf Steven Duplij} \thanks{Alexander von Humboldt Fellow}
\thanks{On leave of absence from {\sl Theory Division, Nuclear Physics
 Laboratory,
Kharkov State University, KHARKOV 310077, Ukraine}}
\thanks{E-mail: duplij@physik.uni-kl.de}\\
{\sl Physics Department, University of Kaiserslautern},\\
{\sl Postfach 3049, D-67653 KAISERSLAUTERN},\\
{\sl Germany}}
\title{{\bf  SOME ABSTRACT PROPERTIES OF SEMIGROUPS APPEARING IN SUPERCONFORMAL
 THEORIES}
}
\date{April 5, 1995}

\newcommand{\limfunc}{\mbox}

\begin{document}

\maketitle
\begin{abstract}
A new type of semigroups which appears while dealing with $N=1$
superconformal symmetry in superstring theories is considered. The ideal series
having
unusual abstract properties is constructed. Various idealisers are
introduced and studied. The ideal quasicharacter is defined. Green's
relations are found and their connection
with the ideal quasicharacter is established.
\end{abstract}
\begin{flushright}KL-TH-95/11
\end{flushright}
\begin{flushleft}hep-th/9505179
\end{flushleft}
\newpage\

\section{Introduction}

Mathematical objects with new properties often appear from concrete physical
considerations and models. The discovery of supersymmetry
\cite{vol/aku,wes/zum2}
gave many new mathematical features, but its influence on the general
abstract properties of the theory, in spite of the fact that among principal
objects there were noninvertible ones and zero divisors \cite{huckaba},
needs to be emphasized. The latter leaded  to the conclusion that
the abstract ground of supersymmetric theory should have semigroup nature
\cite{dup6}. It was also realised that the noninvertible transformations
and semigroups appearing in that way
have many new nontrivial  properties \cite{dup7,dup10}. In particular, it
would be interesting to work out the general abstract structure of the $N=1$
superconformal semigroup, which is important in  the consistent construction of
the
superstring unified theories \cite{kaku1,gre/sch/wit}. In this paper we
provide a consideration of the superconformal semigroups from
the abstract-algebraic point of view and present
their abstract properties without proofs which will appear elsewhere.

\section{Preliminaries}

The semigroup of $N=1$ superconformal transformations of $C^{1,1}$ complex
superspace with the coordinates $(z,\theta )$ valued in the Grassmann
algebra \cite{berezin}, where $z\in C^{1,0}$ and $\theta \in C^{0,1}$, is
isomorphic to the semigroup ${\bf S\ }$ of the even $C^{1,0}\rightarrow
C^{1,0}$ and odd $C^{1,0}\rightarrow C^{0,1}$ functions satisfying some
multiplication law (for details see \cite{dup7,dup10}). The even part of the
law
\begin{equation}
\label{0}{\bf s}_3{\bf =s}_1{\bf *s}_2{\bf ,\ \ s}_i{\bf \in S,\ }
\end{equation}

\noindent in terms of the even functions $g(z)$ can be presented as

\begin{equation}
\label{1}g_3(z)=[g_1(\widetilde{z})+h_1(\widetilde{z})]\cdot g_2(z),
\end{equation}
\noindent where $\widetilde{z}$ is some shifting and $h_1(z)$ is some even
nilpotent function of second degree, i.e. $h_1^2(z)\equiv h_1(z)\cdot
h_1(z)=0$. We stress that, because of the shifting $z\rightarrow \widetilde{z%
}$ and the second term in the brackets (\ref{1}), ${\bf S}$ differs from the
semigroups of functions with point by point multiplication \cite{csa/thu} as
well, as from the semigroups of functions \cite{mag13,mag6}. This leads to
new unusual abstract properties of ${\bf S}$ considered below. Further we
note that to study this properties it is sufficient to know the formal
expression (\ref{1}) only. This parametrisation of $N=1$ superconformal
transformations was given in \cite{dup7,dup10} (where one can also find the
exact formulas and the concrete background). For other considerations we
refer to \cite{bar/fro/sch1,rab3,rog3,hod3}.

Here we do not consider the physical interpretations of $g(z)$ (see
\cite{bar/fro/sch1,cra/rab}) and stress only that $g(z)$
 controls  invertibility of the
superconformal transformations \cite{dup6}.
 Therefore, the index of $g(z)$ which is defined by

\begin{equation}
\label{2}\limfunc{ind}\,g(z)\stackrel{def}{=}\left\{ n\in Z\mid
g^n(z)=0,g^{n-1}(z)\neq 0\right\}
\end{equation}

\noindent plays a crucial part in the following. We mention here that in (%
\ref{1}) and (\ref{2}) the multiplication is a point by point one in the
Grassmann algebra \cite{berezin} (for clarity sometimes we use a point for
it), but the star in (\ref{0}) denotes the semigroup multiplication.

So the semigroup ${\bf S\ }$ can be divided into two disjoint parts ${\bf %
S=G\cup T,}$ ${\bf \ G\cap T=\emptyset }$, where
\begin{equation}
\label{2a}{\bf G}\stackrel{def}{=}\left\{ {\bf s\in S|\,}\limfunc{ind}%
\,g(z)=\infty \right\} ,
\end{equation}
\begin{equation}
\label{2b}{\bf T}\stackrel{def}{=}\left\{ {\bf s\in S|\,}\limfunc{ind}%
\,g(z)<\infty \right\} .
\end{equation}
Here ${\bf G\ }$is a group corresponding to the invertible
transformations. From the multiplication law (\ref{1}) it follows that ${\bf %
T}$ is a two-sided ideal. The unity element ${\bf e\in S\ }$has $g(z)=1,\
h(z)=0,$ and the zero element has $g(z)=0,\ h(z)=0$ (for other details see $%
\cite{dup6,dup10}).$ From (\ref{1}) and the relation $\limfunc{ind}\,h(z)=2$
it follows that ${\bf T\ }$is a nilsemigroup \cite{lal,gri1,gar,sul1}, i.e. $%
\forall {\bf t\in T\ \ \exists }n\in Z,\ {\bf t}^{*n}{\bf =z}$ (here the
multiplication in the power expression is implied as the semigroup one (\ref
{0}))${\bf .\ }$So every element from ${\bf T\ }$is nilpotent without bound
on its index and of finite order, but every element from ${\bf G}$ is of
infinite order.

The superconformal transformations corresponding to  ${\bf G\ }
$ were studied earlier in \cite{bar/fro/sch1,cra/rab,ros/sch/vor1}. Therefore
we
concentrate our attention on the ideal  ${\bf T\ }$, which gives the
evidence of some unusual abstract properties of such parametrised
 superconformal semigroup ${\bf %
S.\ }$

\section{Ideal series}

To classify the elements from the ideal part ${\bf T\ }$ we take $n$-th
power of the equation (\ref{1}) in the Grassmann algebra and, using the
relation $\limfunc{ind}\,h(z)=2$, obtain
\begin{equation}
\label{3}g_3^n(z)=\left[ g_1^n(\tilde z)+n\cdot g_1^{n-1}(\tilde z)\cdot
h_1(\tilde z)\right] \cdot g_2^n(z).
\end{equation}

We see that the natural classification can be done by means of the index of $%
g(z)$ (see (\ref{2})). Let us define the following sets

\begin{equation}
\label{4}{\bf \Delta }_n\ \stackrel{def}{=}\left\{ {\bf s\in S|}\limfunc{ind}%
\,g(z)=n\right\} .
\end{equation}

\begin{equation}
\label{5}{\bf I}_n\ \stackrel{def}{=}\bigcup_{k\leq n}{\bf \Delta }_k.\
\end{equation}

Then we notice that ${\bf T}$ is a disjoint union of the sets ${\bf \Delta }%
_n$, because ${\bf T}=\bigcup_n{\bf \Delta }_n,\ \ {\bf \Delta }_n\cap {\bf %
\Delta }_{n-1}=\emptyset .$ From (\ref{3}) it follows that ${\bf I}%
_{n-1}\subset {\bf I}_n$ and ${\bf I}_n\setminus {\bf I}_{n-1}={\bf \Delta }%
_n.$ Therefore we obtain the following infinite chain of the sets ${\bf I}_n$%
\begin{equation}
\label{6}{\bf z\subset I}_1\subset {\bf I}_2\subset \ldots \subset {\bf I}%
_n\subset \ldots \subset {\bf T.}
\end{equation}

To understand the meaning of ${\bf I}_n$ we use (\ref{3}) and obtain
\begin{equation}
\label{7a}
\begin{array}{lll}
{\bf S*I}_n & \subseteq  & {\bf I}_n
\end{array}
,
\end{equation}
\begin{equation}
\label{7b}
\begin{array}{lll}
{\bf I}_n*{\bf S} & \subseteq  & {\bf I}_{n+1}
\end{array}
,
\end{equation}
\begin{equation}
\label{7c}
\begin{array}{lll}
{\bf S*I}_n*{\bf S} & \subseteq  & {\bf I}_{n+1}
\end{array}
{}.
\end{equation}
{}From these relations we can easily observe that the sets ${\bf I}_n$ are
left ideals of the semigroup ${\bf S}$, but not right ideals, because of (%
\ref{7b}). Moreover, the appearance of $n+1$ in the right side of (\ref{7b})
and (\ref{7c}) is very unusual, and so these strange sets ${\bf I}_n$ is
natural to call ''jumping ideals''. Therefore ${\bf I}_{n-1}\lhd _l{\bf I}_n$
and the chain (\ref{6}) is a left and ''jumping'' ideal series. Then ${\bf I}%
_n$ are quasiideals \cite{steinfeld,cli2} since they satisfy ${\bf S*I}%
_n\cap {\bf I}_n*{\bf S\subseteq I}_n$ . Simultaneously, the sets ${\bf I}_n$
are biideals, because ${\bf I}_n*{\bf S*I}_n\subseteq {\bf I}_n$ \cite
{cat1,laj2}. It is exciting that in our case the regularity is not necessary
for coincidence quasiideals and biideals in superconformal semigroup (as
distinct from \cite{laj2}).Because of the inclusion ${\bf I}_n\lhd {\bf %
U\Rightarrow I}_n\lhd {\bf S,\ \forall U\subset S}$ the semigroup ${\bf S}$
is a filial semigroup \cite{vas}. The indices in (\ref{6}) form a well
ordered set for which $n$ is an ordinal.  Because of ${\bf I}_{n-1}\lhd _l%
{\bf I}_n$ the chain (\ref{6}) can be called a left ascending ideal series
of ${\bf S}$ . From (\ref{7b}) and (\ref{7c}) we derive
\begin{equation}
\label{7d}{\bf S*I}_n\cup {\bf I}_n*{\bf S\subseteq I}_{n+1},
\end{equation}
This condition is opposite for the chain (\ref{6}) to be an ascending
annihilator series of ${\bf S}$ \cite{hme,she3}. So we call it an ascending
antiannihilator series of ${\bf S}$ .

The multiplication law for the sets ${\bf I}_n$ and ${\bf \Delta }_n$ is
\begin{equation}
\label{8}
\begin{array}{lll}
{\bf I}_n*{\bf I}_{n+k} & \subseteq  & {\bf I}_{n+1}, \\ {\bf I}_{n+k-1}*%
{\bf I}_n & \subseteq  & {\bf I}_n, \\ {\bf \Delta }_n*{\bf \Delta }_{n+k} &
\subseteq  & {\bf I}_{n+1}, \\ {\bf \Delta }_{n+k-1}*{\bf \Delta }_n &
\subseteq  & {\bf I}_n, \\ {\bf I}_n*{\bf \Delta }_{n+k} & \subseteq  & {\bf %
I}_{n+1}, \\ {\bf I}_{n+k-1}*{\bf \Delta }_n & \subseteq  & {\bf I}_n, \\
{\bf \Delta }_n*{\bf I}_{n+k} & \subseteq  & {\bf I}_{n+1}, \\ {\bf \Delta }%
_{n+k-1}*{\bf I}_n & \subseteq  & {\bf I}_n, \\ {\bf I}_n*{\bf G} &
\subseteq  & {\bf I}_{n+1}, \\ {\bf G*I}_n & \subseteq  & {\bf I}_n, \\ {\bf %
\Delta }_n*{\bf G} & \subseteq  & {\bf I}_{n+1}, \\ {\bf G*\Delta }_n &
\subseteq  & {\bf \Delta}_n.
\end{array}
\end{equation}
where $k>0$ . It follows that the set ${\bf I}_n$ is a subsemigroup of ${\bf %
S}$ , because from (\ref{8}) we have ${\bf I}_n*{\bf I}_n\subseteq {\bf I}_n$
but the set ${\bf \Delta }_n$ is not a subsemigroup, since ${\bf \Delta }_n*%
{\bf \Delta }_n\subseteq {\bf I}_n$. This is a consequence of the fact that
our semigroup is defined over the Grassmann algebra \cite{berezin} which
contains nilpotents and zero divisors, and the latter fact should be taken
into account properly \cite{huckaba}.

{}From the last two relations of (\ref{8}) and (\ref{7c}) we can obtain
\begin{equation}
\label{9}{\bf G*\Delta }_n*{\bf G\subseteq I}_{n+1},
\end{equation}
i.e. some of the elements from ${\bf \Delta }_n$ are conjugated by the
subgroup ${\bf G}$ with the elements of the next set ${\bf \Delta }_{n+1}$.
By analogy with \cite{sym,lev/wil,lev2} we define $G$-normal subsets ${\bf %
A,B\subseteq S}$ as follows ${\bf g^{-1}*A*g\subseteq B,\ \,g\in G}$. Then
from (\ref{9}) we make a conclusion that any two sets ${\bf \Delta }_n$
contain $G$-normal elements and one can reach any ${\bf \Delta }_n$ using
the subgroup action only. Further general abstract properties of such
elements can be found in \cite{lev/wil,sche3}.

\section{Idealisers}

The left (right, two-sided) idealiser $I_l\,{\bf \left( U\right) }$ ($I_r\,%
{\bf \left( U\right) }$, $\,I\,{\bf \left( U\right) }$)${\bf \ }$of the
subset ${\bf U\subseteq S\ }$ can be defined as the largest subsemigroup of $%
{\bf S\ }$ within which ${\bf U}$ is a left (right, two-sided) ideal, i.e.
\begin{equation}
\label{10a}I_l\,{\bf \left( U\right) }\stackrel{def}{=}\left\{ {\bf %
s\subseteq S\ |\ s*U\subseteq U}\right\} ,
\end{equation}
\begin{equation}
\label{10b}I_r\,{\bf \left( U\right) }\stackrel{def}{=}\left\{ {\bf %
s\subseteq S\ |\ U*s\subseteq U}\right\} ,
\end{equation}
\begin{equation}
\label{10c}I\,{\bf \left( U\right) }\stackrel{def}{=}\left\{ {\bf s\subseteq
S\ |\ s*U\subseteq U,\ U*s\subseteq U}\right\} .
\end{equation}

For the set $I\,({\bf U)}$ the set ${\bf U}$ is really a subsemigroup,
because ${\bf U*s\subseteq U}$, ${\bf s*U\subseteq U}$, ${\bf U*t\subseteq U}
$, ${\bf t*U\subseteq U}\Rightarrow {\bf U*s*t}\subseteq {\bf U*t\subseteq U}
$, ${\bf s*t*U}\subseteq {\bf s*U\subseteq U}$. Also, if {\bf V} is a
subsemigroup of ${\bf U}$ and {\bf V}$\lhd {\bf U\ }$, then $\forall {\bf %
v\in V\Rightarrow }$ ${\bf v*U\subseteq U}$, ${\bf U*v\subseteq U\Rightarrow
v}\in I\,\left( {\bf U}\right) $. Thus ${\bf V}\subseteq I\,\left( {\bf U}%
\right) .$

Let us consider the idealisers of the various introduced subsets of ${\bf S}$
. First the left idealiser for ${\bf I}_n$ is ${\bf S}$, as is follows
directly from (\ref{7a}), i.e.
\begin{equation}
\label{11}I_l\,\left( {\bf I}_n\right) ={\bf S.}
\end{equation}

{}From the last relation in (\ref{8}) we find
\begin{equation}
\label{11a}I_l\,\left( {\bf \Delta }_n\right) ={\bf G.}
\end{equation}

For the right idealisers of ${\bf I}_n$ the situation is more
complicated.Using (\ref{7b}) we divide ${\bf S}$ into two disjoint parts $%
{\bf S=S^I}_n\cup {\bf S^\Delta }_n$, where ${\bf S^I}_n\cap {\bf S^\Delta }%
_n=\emptyset $, and they satisfy the relations
\begin{equation}
\label{12a}
\begin{array}{lll}
{\bf I}_n*{\bf S^I}_n & \subseteq & {\bf I}_n
\end{array}
,
\end{equation}
\begin{equation}
\label{12b}
\begin{array}{lll}
{\bf I}_n*{\bf S^\Delta }_n & \subseteq & {\bf \Delta }_{n+1}
\end{array}
{}.
\end{equation}

By definition (\ref{10b}) ${\bf S^I}_n$ is the right idealiser for ${\bf I}%
_n $, i.e.
\begin{equation}
\label{13}I_r\,\left( {\bf I}_n\right) ={\bf S^I}_n.
\end{equation}

Obviously, that ${\bf I}_n\subset {\bf S^I}_n$, since ${\bf I}_n*{\bf I}%
_n\subset {\bf I}_n$. Therefore ${\bf S^I}_n={\bf I}_n\cup {\bf S^{II}}_n$.
{}From (\ref{3}) it follows that for the elements from ${\bf S^{II}}_n$ the
second term in the brackets should disappear, therefore we find
\begin{equation}
\label{14}{\bf S^{II}}_n=\left\{ {\bf s\in T\setminus I}_n|\ %
g_1^{n-1}\left( \widetilde{z}\right) \cdot g_2^n\left( z\right) =0, \, %
h_1\left( \widetilde{z}\right) \cdot g_2^n\left( z\right) =0\right\} .
\end{equation}

Then the ''jumping'' set ${\bf S^\Delta }_n$ from (\ref{12b}) is equal to $%
{\bf S^\Delta }_n=\left( {\bf S\setminus I}_n\right) \setminus {\bf S^{II}}%
_n $.

Another way to vanish the second term in (\ref{3}) is the consideration of
the special superconformal transformations (they are called
{\it Ann-}transforma\-tions in \cite{dup10}) for which the relation
$g^{n-1}\left( z\right)
\cdot h\left( z\right) =0$ is valid (see (\ref{1}) and (\ref{3})). Let us
divide ${\bf I}_n$ in two disjoint parts ${\bf I}_n={\bf I^A}_n\cup {\bf %
I^{\neq A}}_n$, where ${\bf I^A}_n\stackrel{def}{=}\left\{ {\bf s\in I}_n|
\,g^{n-1}\left(
z\right) \cdot h\left( z\right) =0\right\} $. It was shown in \cite{dup10}
that {\it Ann}-property is preserved from the right only, and so we obtain $%
{\bf I^A}_n*{\bf S\subseteq I^A}_n$, which means that ${\bf I^A}_n$ is a
right ideal in ${\bf S}$, then
\begin{equation}
\label{14a}I_r\,\left( {\bf I^A}_n\right) ={\bf S.}
\end{equation}

For the sets  ${\bf \Delta }$${\bf ^A}_n={\bf I^A}_n\setminus {\bf I^A}_{n-1}
$ we find ${\bf \Delta ^A}_n*{\bf G\subseteq {\bf \Delta ^A}}_n$, therefore
\begin{equation}
\label{14b}I_r\,\left( {\bf \Delta ^A}_n\right) ={\bf G.}
\end{equation}

We note here that by means of the right group action we can reach a set $%
{\bf I}_n$ with any large $n$, because the relation ${\bf \Delta ^{\neq A}}%
_n*{\bf G\subseteq {\bf \Delta ^{\neq A}}}_{n+1}$ (see also (\ref{9})).

\section{Ideal quasicharacter}

Let us define
\begin{equation}
\label{15}\chi \left( {\bf s}\right) \stackrel{def}{=}\left\{ n\in N|\,
\limfunc{ind}\,g\left( z\right) =n\right\} .
\end{equation}

Using (\ref{7a}) and (\ref{7b}) we obtain
\begin{equation}
\label{16}\limfunc{max}\,\chi \left( {\bf s*t}\right) = \left\{
\begin{array}{cccc}
\chi \left( {\bf t}\right) , \, & \chi \left( {\bf s}\right) & \geq &
\chi \left(
{\bf t}\right) \\ \chi \left( {\bf s}\right) +1, & \chi \left( {\bf s}%
\right) & < & \chi \left( {\bf t}\right) .
\end{array} \right.
\end{equation}

In particular,
\begin{equation}
\label{17}
\begin{array}{ccc}
\chi \left( {\bf g*s}\right) & = & \chi \left(
{\bf s}\right) , \\ \chi \left( {\bf s*g}\right) & = & \chi \left( {\bf s}%
\right) +1.
\end{array}
\end{equation}

{}From (\ref{16}) it follows that  $n_\delta =\left| \chi \left( {\bf s*t}%
\right) -\chi \left( {\bf s}\right) -\chi \left( {\bf t}\right) \right| $ is
bounded. This value $n_\delta $ shows how much the mapping ${\bf s}%
\rightarrow \chi \left( {\bf s}\right) $ differs from a homomorphism
\cite{kaz}%
. The limitedness of $n_\delta $ allows us to conclude that $\chi \left( {\bf
s%
}\right) $ is a quasicharacter \cite{sht1} which can be called an ideal
quasicharacter. The elements of ${\bf S}$ having finite ideal quasicharacter
are nilpotent and belong to the ideal ${\bf T}$, and $\chi \left( {\bf g}%
\right) =\infty ,\,{\bf g\in G}$. Another description of the ideal
quasicharacter can be written as follows $\chi \left( {\bf s}\right)
=n\Longleftrightarrow {\bf s\in \Delta }_n$. Since ${\bf \Delta }$$_n\cap
{\bf \Delta }_m=\emptyset ,\,n\neq m$ , we conclude that $\chi \left( {\bf s}%
\right) $ indeed disjoins the elements of ${\bf S}$, and the relation $\pi $
defined as {\bf s$\pi t\Longleftrightarrow $}$\chi \left( {\bf s}\right)
=\chi \left( {\bf t}\right) $ is an equivalence relation in ${\bf S}$.

\section{Green's relations}

In our notations the Green's ${\cal L}$ and ${\cal R}$ relations are
\begin{equation}
\label{18}
\begin{array}{ccccc}
{\bf s\ }{\cal L\ }{\bf t} & \Longleftrightarrow & \exists {\bf u,v\in S,} &
{\bf u*s=t,} & {\bf v*t=s,} \\ {\bf s\ }{\cal R\ }{\bf t} &
\Longleftrightarrow & \exists {\bf u,v\in S,} & {\bf s*u=t,} & {\bf t*v=s.}
\end{array}
\end{equation}

Let us find ${\cal L}$ and ${\cal R}$ equivalent elements in the
superconformal semigroup ${\bf S}$. Using (\ref{7a}) and (\ref{16}) we find
that ${\bf s\ }{\cal L\ }{\bf t\,}\Rightarrow \chi \left( {\bf s}\right)
\leq \chi \left( {\bf t}\right) \wedge \chi \left( {\bf t}\right) \leq \chi
\left( {\bf s}\right) \Rightarrow \chi \left( {\bf s}\right) =\chi \left(
{\bf t}\right) $. Therefore ${\cal L}=\pi $, and ${\cal L}$-equivalent
elements have the same ideal quasicharacter,
\begin{equation}
\label{19}{\bf s\ }{\cal L\ }{\bf t\Rightarrow }\chi \left( {\bf s}\right)
=\chi \left( {\bf t}\right) ,
\end{equation}
and they belong to the same set ${\bf \Delta }_n$. By analogy from (\ref{7b}%
) for the ${\cal R}$-equivalent elements we derive ${\bf s\ }{\cal R\ }{\bf %
t\,}\Rightarrow \chi \left( {\bf s}\right) \leq \chi \left( {\bf t}\right)
+1\wedge \chi \left( {\bf t}\right) \leq \chi \left( {\bf s}\right) +1$.
Then the ideal quasicharacters of the ${\cal R}$-equivalent elements can
differ only by 1 or coincide, i.e.
\begin{equation}
\label{20}{\bf s\ }{\cal R\ }{\bf t\,\Rightarrow \left| \chi \left( s\right)
-\chi \left( t\right) \right| \leq 1.}
\end{equation}

Since ${\cal H=L\cap R}$, the sets ${\bf \Delta }_n$ consist also of ${\cal H%
}$-equivalent elements.

Consider the ${\cal L}$-equivalent elements. Let ${\bf s\neq t,s\neq z,t\neq z%
}$. From (\ref{18}) we derive that ${\bf s=v*\left( u*s\right) =\left(
v*u\right) *s=\,\left( v*u\right) }^{*k}*{\bf s}$ for any $k\in N$. If ${\bf %
v\in T\vee u\in T}$, then ${\bf \left( v*u\right) }^{*k}\in {\bf T}$, since%
{\bf \ }${\bf T}${\bf \ }is an ideal in ${\bf S}$. Because of ${\bf T}$ is a
nilsemigroup $\exists n\in N$ such that ${\bf \left( v*u\right) }^{*n}={\bf z%
}$. Through the arbitrariness of $k$ we choose $k=n$ and obtain ${\bf %
s=\left( v*u\right) }^{*n}*{\bf s=z*s=z}$ or ${\bf s=t}$, which contradicts
the initial assumptions. The same is valid for other Green's relations.
Therefore ${\bf v\in G\wedge u\in G}$, i.e. nontrivial ${\cal L}$ and ${\cal %
R}$ equivalences can be constructed with regard to the invertible elements
of ${\bf S}$ only. Then the principal left and right ideals generated by $%
\forall $${\bf t\in S}$ and defined by ${\bf L\left( t\right) \,}\stackrel{%
def}{=}{\bf S*t}$ and ${\bf R\left( t\right) }\stackrel{def}{=}{\bf t*S}$,
as a matter of fact are some analogies of the left and right cosets of ${\bf %
G}$ in ${\bf S}$ introduced in \cite{mca3,sche3}.
\section{Acknowledgements}
The author is grateful to Prof. J. M. Howie for fruitful conversations and
remarks and
the kind hospitality at
the University of
  St. Andrews, where the work was begun. Also the
discussions with D. A. Arinkin, V. G. Drinfeld, J. Kupsch, M.
V. Lawson, B. V. Novikov, W. R\"uhl, S. D. Sinelshchikov and J.~Wess are
greatly acknowledged.

\end{document}